\pdfoutput=1
\documentclass[a4paper,10pt]{article}
\usepackage[english]{babel}
\usepackage[utf8]{inputenc}

\usepackage{amsmath}
\usepackage{graphicx}
\usepackage{url}

\title{Speaker Diarization: Using Recurrent Neural Networks}

\author{{Vishal Sharma, Zekun Zhang, Zachary Neubert, Curtis Dyreson} \\ {Utah State University} \\ {Logan, Utah}}

\begin{document}
\maketitle

\begin{abstract}
Speaker Diarization is the problem of separating speakers in an audio. There could be any number of speakers and final result should state when speaker starts and ends. In this project, we analyze given audio file with 2 channels and 2 speakers (on separate channel). We train Neural Network for learning when a person is speaking. We use different type of Neural Networks specifically, Single Layer Perceptron (SLP), Multi Layer Perceptron (MLP), Recurrent Neural Network (RNN) and Convolution Neural Network (CNN) we achieve  $\sim$92\% of accuracy with RNN. The code for this project is available at \url{https://github.com/vishalshar/SpeakerDiarization_RNN_CNN_LSTM}. 
\end{abstract}


\medskip
\section{Introduction}

In this section, we introduce what the original problem is, how we formulate it into a neural network problem, and what criteria we use to evaluate the performance of neural network. The benchmark is also provided given our evaluation criteria.

Speaker diarization is the task of determining "who spoke when?"\cite{anguera2010speaker} in an audio with unknown number of speakers and unknown length of audio. It is of great importance in the domain of speech research and has several applications in the field of information retrieval \cite{anguera2012speaker} \cite{pawelDeepRNN}.

\subsection{Original Problem}
The provided data set is a collection of audio files (“.wav” files) along with the corresponding labels (“.csv” files) indicating the time period of each speech from each speaker. The task is to train a neural network which can generate the diarization results automatically given an audio file.

The provided audio files are “.wav” files with sample rate 44100. The content of each audio file is a conversation between speakers. Each audio file has two channels, one for each speaker. More specifically, in each audio file, channel 1 is recorded by the microphone in front of speaker 1 and channel e2 is recorded by the microphone in front of speaker 2.

The provided labels are “.csv” files. There are multiple types of “.csv” files provided, the one we used throughout this project is like below

\begin{center}
	\begin{tabular}{ c c c c }
		tmin & tier & text & tmax \\ 
		0	&	CH2&	N&	1.361079 \\
		0	&	CH1&	N&	4.550097 \\
		1.361079&	CH2&	S&	4.996529 \\
		4.550097&	CH1&	S&	5.541915 \\ 
		4.996529&	CH2&	N&	5.547973 \\
		5.541915&	CH1&	N&	8.183008 \\
	\end{tabular}
\end{center}

\subsection{Formulated Problem}

We formulate this task to a supervised classification problem. Generally, we divide each audio file into small segments on time domain with fixed segment size and use trained neural network to classify whether a segment belongs to the speech of a speaker. The segment size is a parameter and can be changed to any number in our code. If the segment size is too small, then the samples within a segment may be not enough to do the classification and hence the performance is poor. If the segment size is too big, then the precision of our final output is low compared to the provided labels. Throughout this project, we set segment size as 0.1 second to achieve a good performance as well as high precision. The original audio file is segmented as the graph shown below. For example, the length of an audio file is 10 seconds and we set the segment size as 0.1 second. The length of each channel is also 10 s. Then each channel can be divided into 100 segments. As the sample rate of provided audio file is 44100, each single segment contains 4410 samples. 

\begin{center}
	\begin{tabular}{|c|c|c|c|c|}
		\hline
		\multicolumn{5}{|c|}{Channel\textunderscore1} \\
		\hline
		Segment\textunderscore1 & Segment\textunderscore2 & Segment\textunderscore3 & Segment\textunderscore4 &  $\dots$\\
		\hline
	\end{tabular}
\end{center}

Given a segment, we predict the class of this segment. There are two types of classifications we tried in this project as below

\begin{itemize}
	\item 2 classes: 
	\subitem 0:Non-speech (including noise, speech of another speaker)
	\subitem 1:Speech of the speaker of the current channel   \\
	\subitem Detail: this type of classification can only be used when treating each channel of an audio file separately. For example, we extract two channels, channel\textunderscore1 and channel\textunderscore2, from an audio file. Then for a segment of channel\textunderscore1, it belongs to class 1 if this segment is part of the speech of speaker 1, and belongs to class 0 otherwise. For a segment of channel\textunderscore2,

	\item 4 classes:
	\subitem 0: Non-speech (including noise, speech of another speaker)
	\subitem 1: Speech of speaker 1
	\subitem 2: Speech of speaker 2
	\subitem 3: Overlap of two speakers \\
	\subitem Detail: This classification can be used for either treating two channels separately or considering them together. However, in practice this kind of classification does not perform as well as the 2 classes method. In this project, we mainly use 2 classes method and will briefly introduce how to implement 4 classes method.
\end{itemize}

\medskip
\section{Dataset}
Our dataset contains 37 audio files approximately of 15 minutes each with sampling rate of 44100 samples/second, recorded in 2 channels with exactly 2 speakers on 2 different microphones. Each audio file has been hand annotated for speakers timings. Annotating timing (in seconds) they start and stop speaking. We use this dataset and split in 3 parts for training, validation and testing.

\subsection{Data Normalization}
We perform normalization of audio files after observing recorded audio was not in the same scale. Few audio files were louder than others and normalization can help bring all audio files to same scale. Normalized audio files are generated by normalizing all original audio files to the same average volume (-20 dBFS). Normalization is implemented by code in “Normalize Audio.py”. At the beginning of this project, we didn’t implement normalization cause we thought all files are at similar level. Later we found that the accuracy can achieve 90\% on file 1-file 30, however, can only get 60\% on file 31-file 37. By listening them, we found file 31-file 37 are not as loud as other files. After normalization this problem is solved and all files can achieve approximately the same accuracy.

\subsection{Load Audio Data}
“.wav” file is loaded by using the “get$\textunderscore$data” function written in “Load$\textunderscore$Audio$\textunderscore$Data.py” which is attached at the end of the report. Specifically “get\textunderscore data” function read the “.wav” file located at certain directory and return a tuple (channel$\textunderscore$1$\textunderscore$matrix, channel$\textunderscore$2$\textunderscore$matrix) where each element is a matrix contains the samples of a channel. Each row of a matrix is the samples within a segment. The shape of each matrix is 
(number$\textunderscore$of$\textunderscore$segments$\textunderscore$each$\textunderscore$channel, number$\textunderscore$of$\textunderscore$samples$\textunderscore$each$\textunderscore$segment). 
For example, if the length of a “.wav” file is 10 seconds, then two matrixes in the returned tuple is as below. Recall that the segment size is 0.1 second, so we have 100 segments hence there are 100 rows in each matrix. Sample rate is 44100 which means there are 44100 samples in one second and 4410 samples in each segment. In the graph below we use index 1.

\begin{center}
	\begin{tabular}{|l|l|}
		\hline
		\multicolumn{1}{|c|}{channel$\textunderscore$1$\textunderscore$matrix} & \multicolumn{1}{|c|}{channel$\textunderscore$2$\textunderscore$matrix} \\ 
		\hline
		Segment 1 (sample 1-4410)	&	Segment 1 (sample 1-4410)    \\
		Segment 2 (sample 4411-8820)	&	Segment 2 (sample 4411-8820)     \\
		Segment 3 (sample 8821-13230) &	Segment 3 (sample 8821-13230) \\
		$\dots$ &	$\dots$  \\ 
		Segment 100 (sample 436591-441000) &	Segment 100 (sample 436591-441000) \\
		\hline
	\end{tabular}
\end{center}

“get data” function can also down sample the original data. For example, if set argument “down$\textunderscore$sample = true” and “down$\textunderscore$sample$\textunderscore$rate =4”, then it only retain the 1st sample of every 4 samples (discarding the residue). For the example above, after down sampling the number of samples in each segment is 1102.  We use “down$\textunderscore$sample$\textunderscore$rate=4” throughout this project

\subsection{Sampling Audio}
With frame rate being high, we have a lot of data. To give an example, in a 15 min audio file we get about 40M samples in each channel.  To reduce data without loosing much information, we down sample audio files by every 4 sample. \cite{giannakopoulos2015pyaudioanalysis}

\subsection{Load Labels}
“.csv” file is loaded by using the “get$\textunderscore$labels” function in “Labels.py”. Specifically “get$\textunderscore$labels” function read the “.csv” file located at certain directory and return a tuple (label$\textunderscore$1, label$\textunderscore$2) where each element is a matrix (actually it is a binary vector) contains the labels for each channel. The shape of each matrix is
(number$\textunderscore$of$\textunderscore$segments$\textunderscore$each$\textunderscore$channel,1)
For example, the returned label matrixes corresponding to the data matrixes above are:

\begin{center}
	\begin{tabular}{|l|l|}
		\hline
		\multicolumn{1}{|c|}{Label$\textunderscore$1$\textunderscore$matrix} & \multicolumn{1}{|c|}{Label$\textunderscore$2$\textunderscore$matrix} \\ 
		\hline
		0 or 1 (label for segment 1)	&	0 or 1 (label for segment 1)   \\
		0 or 1 (label for segment 2)	&	0 or 1 (label for segment 2)   \\
		0 or 1 (label for segment 3)    &	0 or 1 (label for segment 3)   \\
		$\dots$ &	$\dots$  \\ 
		0 or 1 (label for segment 100)  &	0 or 1 (label for segment 100) \\
		\hline
	\end{tabular}
\end{center}
“get$\textunderscore$labels” and “get$\textunderscore$data” can be override to modify the format of input and labels for different neural networks

We depict how to convert the time period in “.csv” file to the matrix above. Given the “.csv” file as below

\begin{center}
	\begin{tabular}{ c c c c }
		tmin & tier & text & tmax \\ 
		0	&	CH2&	N&	1.361079 \\
		0	&	CH1&	N&	4.550097 \\
		1.361079&	CH2&	S&	4.996529 \\
		4.550097&	CH1&	S&	5.541915 \\ 
		4.996529&	CH2&	N&	5.547973 \\
		5.541915&	CH1&	N&	8.183008 \\
	\end{tabular}
\end{center}

We at first truncate the time by discarding numbers after 0.1, and divide the truncated numbers by segment size 0.1, then the original time is transferred into index as below

\begin{center}
	\begin{tabular}{ c c c c }
		tmin & tier & text & tmax \\ 
		0	&	CH2&	N&	13 \\
		0	&	CH1&	N&	45 \\
		13&	CH2&	S&	49 \\
		45&	CH1&	S&	55 \\ 
		49&	CH2&	N&	55 \\
		55&	CH1&	N&	81 \\
	\end{tabular}
\end{center}

After that we create an all zeros vector for each channel and change the positions where is the speech of corresponding speaker to 1. For example, if channel one has 100 segment as in previous examples, then we create an all zeros vector “label$\textunderscore$1”. As indicated by the row 4 of the table above, index 45-55 is the speech of speaker 1, so we set label$\textunderscore$1[45:55] =1. Labels are generated in this way by reading the whole table.

\subsection{Cleaning Labels}
Provided labels needed some cleaning described below:
\begin{enumerate}
\item Names of the speakers was not consistent throughout the data file, we cleaned it and made sure name is consistent.
\item File also contained unicode, which needed to be cleaned. Python goes crazy with unicodes lol
\item There were miss alignments as well in the data and needed to be removed and fixed.
\end{enumerate}

\subsection{Grouping Data}

There are 37 pairs of “.wav” and “.csv” files in total. We randomly divide them into training set, cross-validation set and testing set with the ratio 70\%/15\%/15\%.

\medskip
\section{Evaluation}
The performance of the network is evaluated by the percentage of correctly classified segments. For example, an audio file in the test set has 100 segments and out trained network can correctly classify 90 out of them, then the accuracy of our classification on this file is 90\%. We use the average accuracy of all the audio files in the test set as the final accuracy of our network. So if there are 3 files in the test set, and the accuracy provided by one network on each file is 88\%, 90\% and 92\%, then we say the accuracy of this network is 90\%.

\subsection{Benchmark}
By analyzing the provided label files (“.csv” files), we find that basically the speech time of a speaker is about 40\%. So if we always guess the majority (non-speech), the benchmark of accuracy is about 60\% for 2 classes method. For 4 classes method, the benchmark is about 30\%.

\medskip
\section{Approach}
In this section we first descripted how we pre-processed data and labels. Then we elaborate all the networks we tried to solve this classification problem.

\subsection{Multi-layer Perceptron}

We first tried a simple multilayer perceptron. The code corresponding to this part of work is in “Alg1 MLP 1channel 2classes”. Each time we feed one row of the loaded data matrix to the network to do classification. Each channel is treated as a single audio file. So the number of input neurons equals to the number of samples in each segment (each row of data matrix). The number of hidden layers and number of neurons there in can be changed to any number. The number of output neurons is just 1 to predict 0 or 1. A typical graph for a 2 layer MLP is shown below

For all the networks used in this project, the hidden neurons are ReLu and the output neuron is sigmoid (for 4 classes method, there are 4 output neurons and the output neuron is softmax). If the value after applying sigmoid is above 0.5, we classify it as 1, otherwise classify it as 0 (in the code we judge whether the logit before sigmoid greater than 0 or not). The cost function is cross entropy. Mini-batch gradient descent with Adam optimization is used to train network. Learning rate is 0.001 (default learning rate) throughout this project. Then number of weights depends on the number of layers and neurons therein. Basically the number of weights is the multiply of number of neurons in every two consecutive layers.

We start with a basic single layer perceptron model. We implement 3 different models with hidden layer of different sizes 100, 200, 500 neurons. We achieve approximately 86\% accuracy. 

We next move to multi-layer perceptron model and try models with 2 layers deep. First layer had 100 and second 50 neurons and another with higher number of neurons (First Layer: 200, Second Layer: 100) (First Layer: 300, Second Layer: 50). For all the networks used in this project, the hidden neurons are ReLu \cite{relu} and the output neuron are sigmoid. The cost function used is cross entropy and mini-batch gradient descent with Adam optimization is used to train network.

\subsection{Recurrent Neural Network (RNN)}
Next we try Recurrent Neural Network \cite{mikolov2010recurrent} on the classification problem. The RNN gives us the best result with 3 layers each with 150 Long short-term memory (LSTM) cells. The LSTM in the graph means a LSTM layer which consists of 150 LSTM cells. The output only has one neuron with sigmoid to predict 0 or 1. 
he code here is in “Alg4 RNN 1channe 2classes”For input, we reshape a segment (one row of data matrix) into a matrix whose shape is 
(number of steps, number of samples each segment// number of steps)
Recall that the number of samples each segment after down sampled is 1102, the number of steps we used is 22, so the shape of input is (22, 50), residue samples are discarded. The number of input neurons is 50, in each time step, one row of reshaped matrix is feed in. After 22 steps, a single output is used to do classification. We tried different parameters (number of layers, number of cells in each layer, number of steps, dropout rate, et al). The RNN giving us the best result has 3 layers with 150 Long short-term memory (LSTM) cells in each layer. The graph of RNN is as below. The LSTM in the graph means a LSTM layer which consists of 150 LSTM cells. The output only has one neuron with sigmoid to predict 0 or 1. We didn’t try 4 classes method on RNN cause the result by using 4 classes method seems not promising as shown in MLP.

\begin{figure}
\centering
\includegraphics[width=0.8\textwidth]{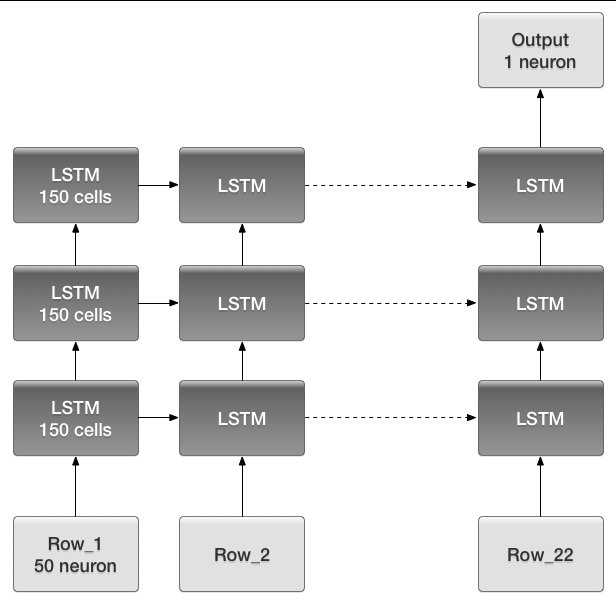}
\caption{\label{fig:RNN}Structure of RNN using TensorFlow.}
\end{figure}

\subsection{Convolution Neural Network}
To apply CNN \cite{krizhevsky2012imagenet}, we at first compute the spectrogram for each row of the data matrix, then store them into a new file by using pickle. In this way we don’t need to compute spectrogram online and hence can save a lot of training time. Function scipy. signal.spectrogram is used to compute the spectrogram for each segment. The recomputed spectrogram of each segment then is organized to a 3 dimension matrix with shape (number of segments, height, width). For example, the down sampled data matrix of a channel returned by get data has the shape (100, 1102) for a channel with 100 segments, then the shape of recomputed spectrogram matrix is (100,129,4). The number of segments remains the same. The height 129 and width 4 come from using the default parameters of function scipy. signal.spectrogram. Spectrogram matrices are computed and stored by using code in Spectrogram Generator.

\begin{figure}
	\centering
	\includegraphics[width=0.5\textwidth]{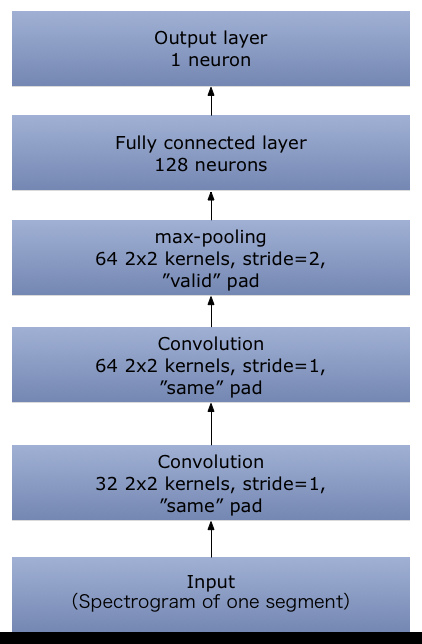}
	\caption{\label{fig:CNN}Structure of CNN using TensorFlow.}
\end{figure}

\subsection{Testing}

Our best result is generated by RNN which achieves 91.47\% on test set. We feed the test data to the network and visualized the result.  The graph below compared our prediction with truth label for a certain period of time of one test file. Left side is the label, right side is our prediction. Vertical line indicates the prediction/label is 1. Horizontal axis is the index of segment, which can be mapped back to time axis. We can see our prediction looks very similar to the labels.
\begin{figure}  
	\centering
	\includegraphics[width=1.0\textwidth]{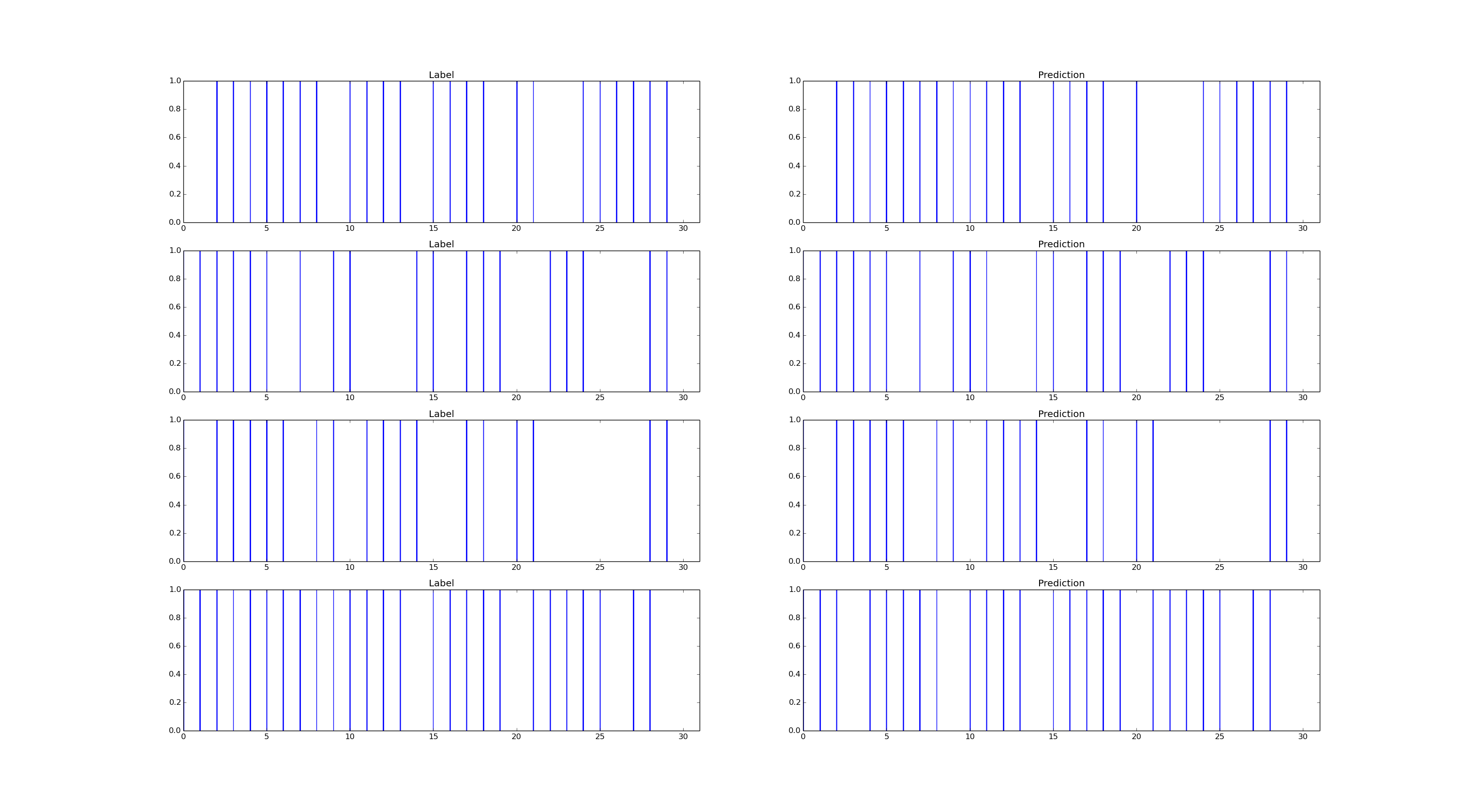}
	\caption{\label{fig:comapre}Compare Predicted with given Label.}
\end{figure}

\begin{figure} 
\centering
\includegraphics[width=1.0\textwidth]{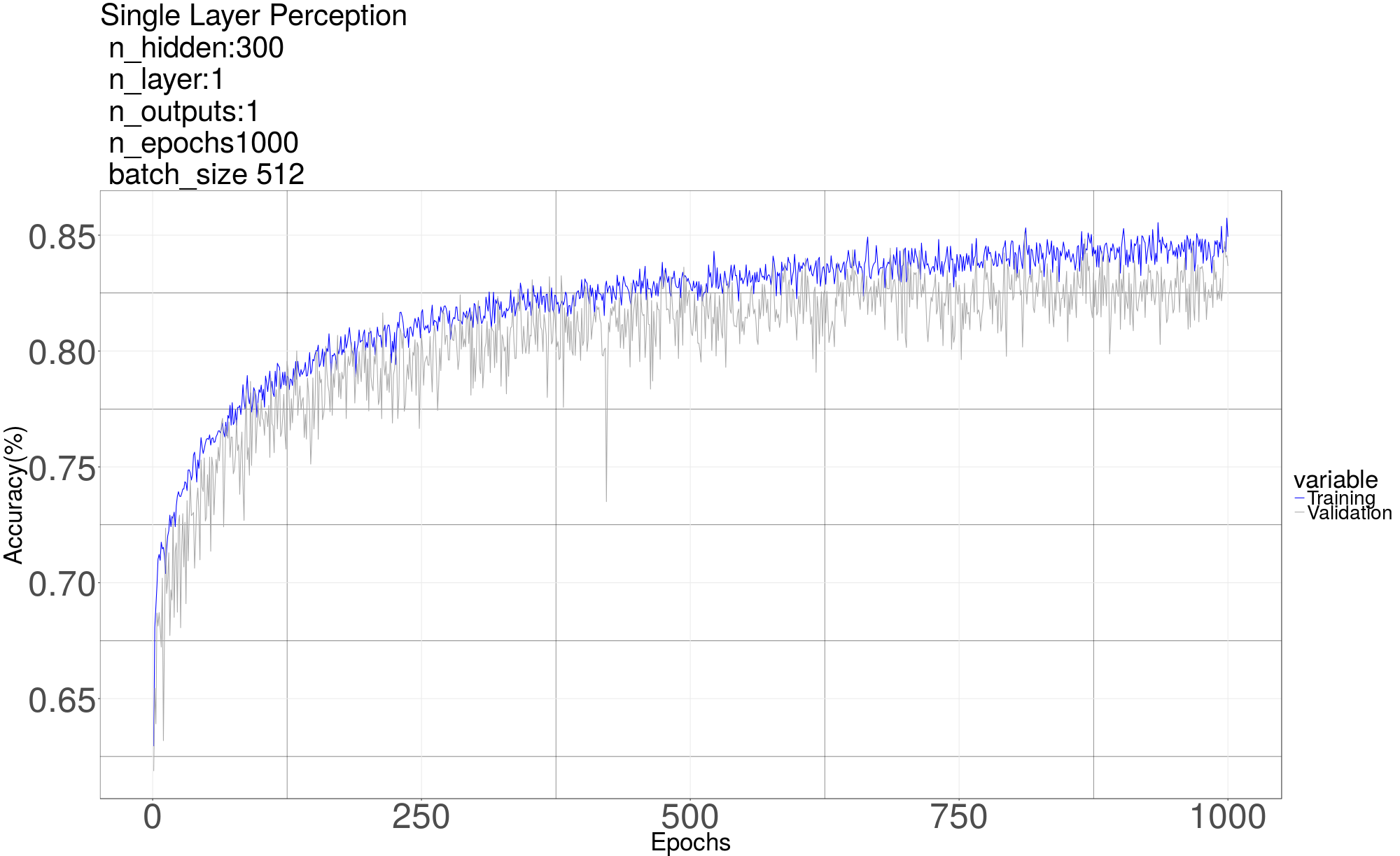}
\caption{\label{fig:MLP_2}Results of Single Layer Perceptron on Training and Validation set. Title contains information about configuration of the network.}
\end{figure}

\begin{figure}  
\centering
\includegraphics[width=1.0\textwidth]{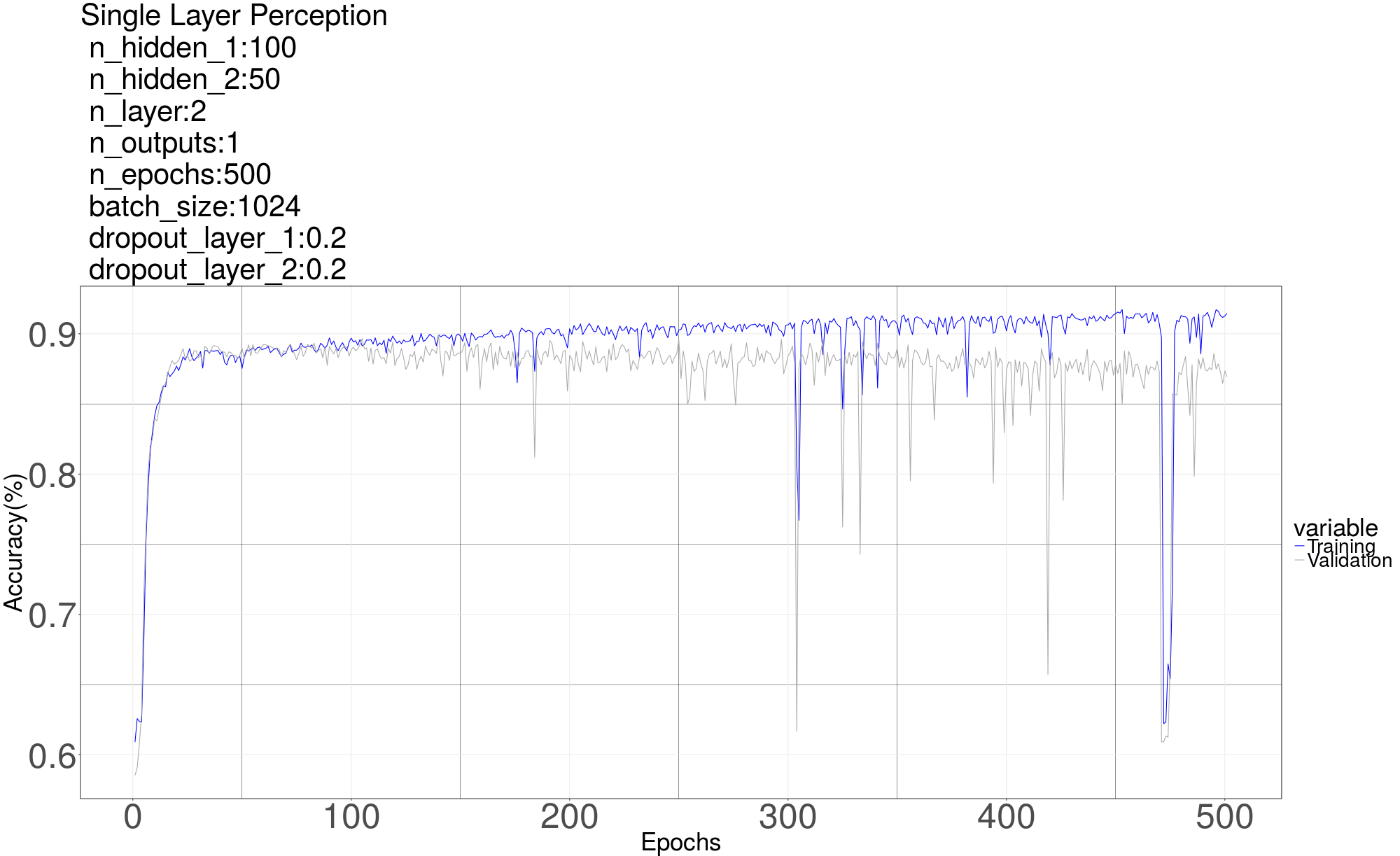}
\caption{\label{fig:MLP_6}Results of Multi Layer Perceptron on Training and Validation set. Title contains information about configuration of the network.}
\end{figure}

\begin{figure}  
\centering
\includegraphics[width=1.0\textwidth]{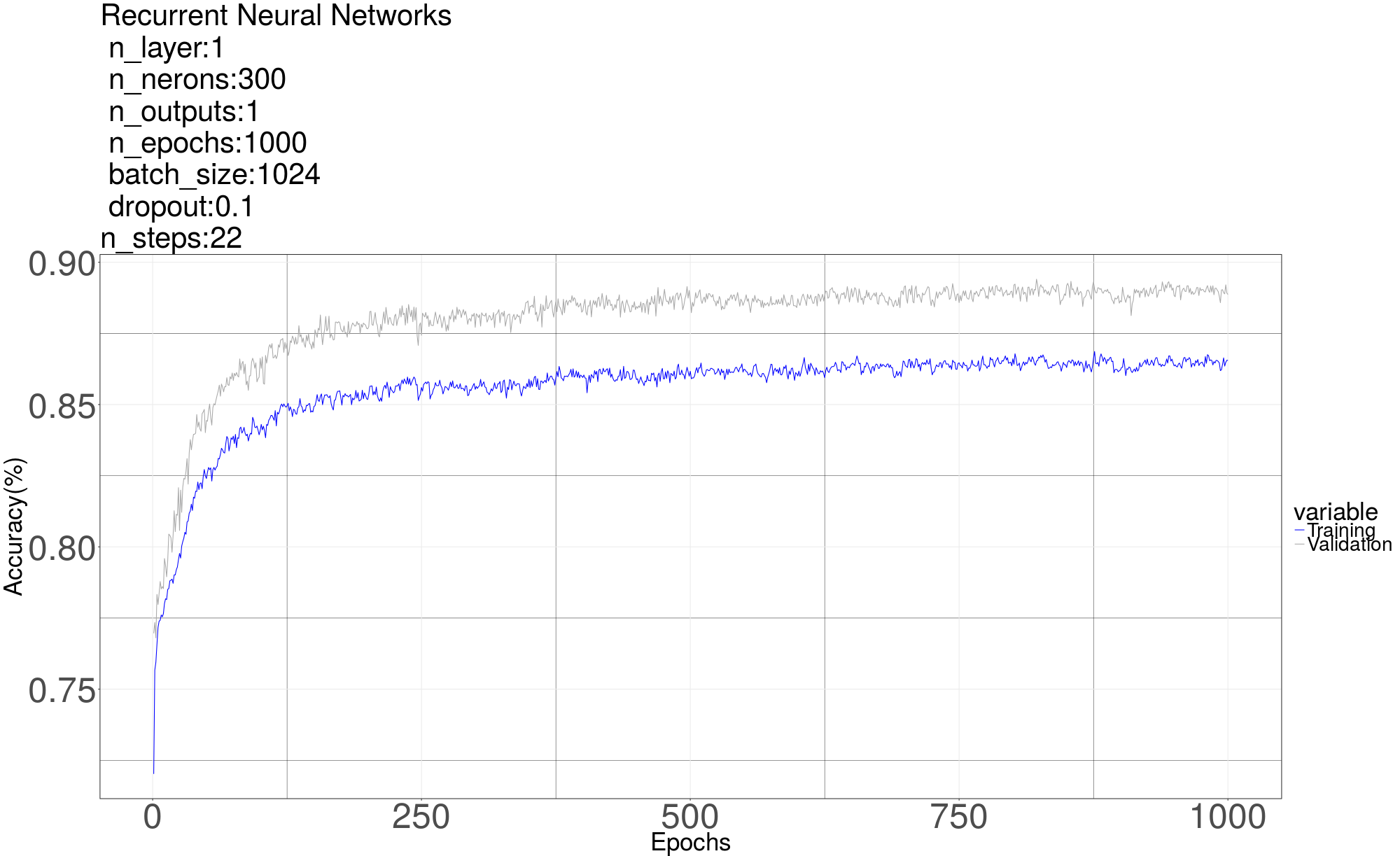}
\caption{\label{fig:RNN_1}Results of Recurrent Neural Network on Training and Validation set. Title contains information about configuration of the network.}
\end{figure}

\begin{figure}  
\centering
\includegraphics[width=1.0\textwidth]{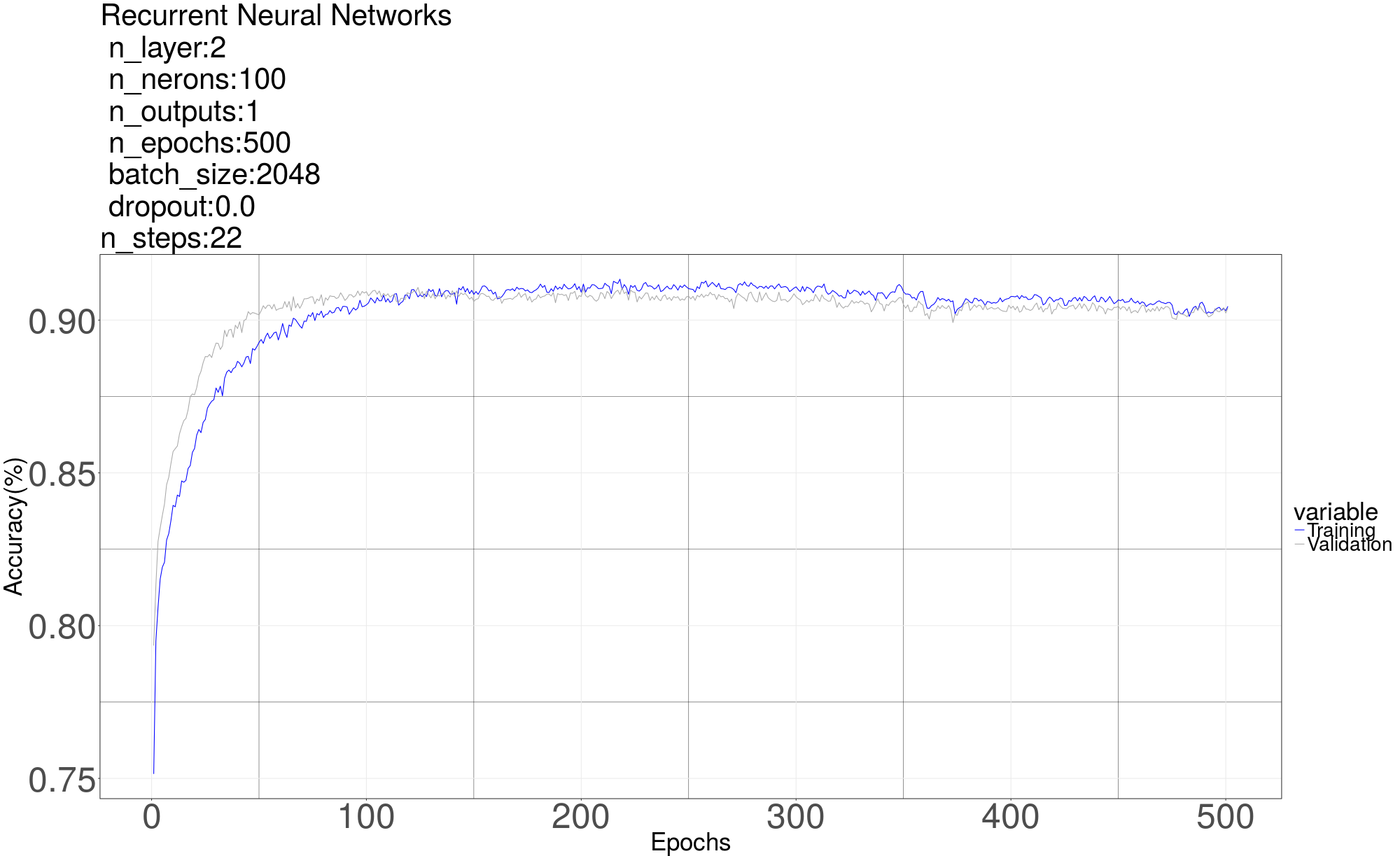}
\caption{\label{fig:RNN_3}Results of Recurrent Neural Network on Training and Validation set. Title contains information about configuration of the network.}
\end{figure}

\begin{figure}  
\centering
\includegraphics[width=1.0\textwidth]{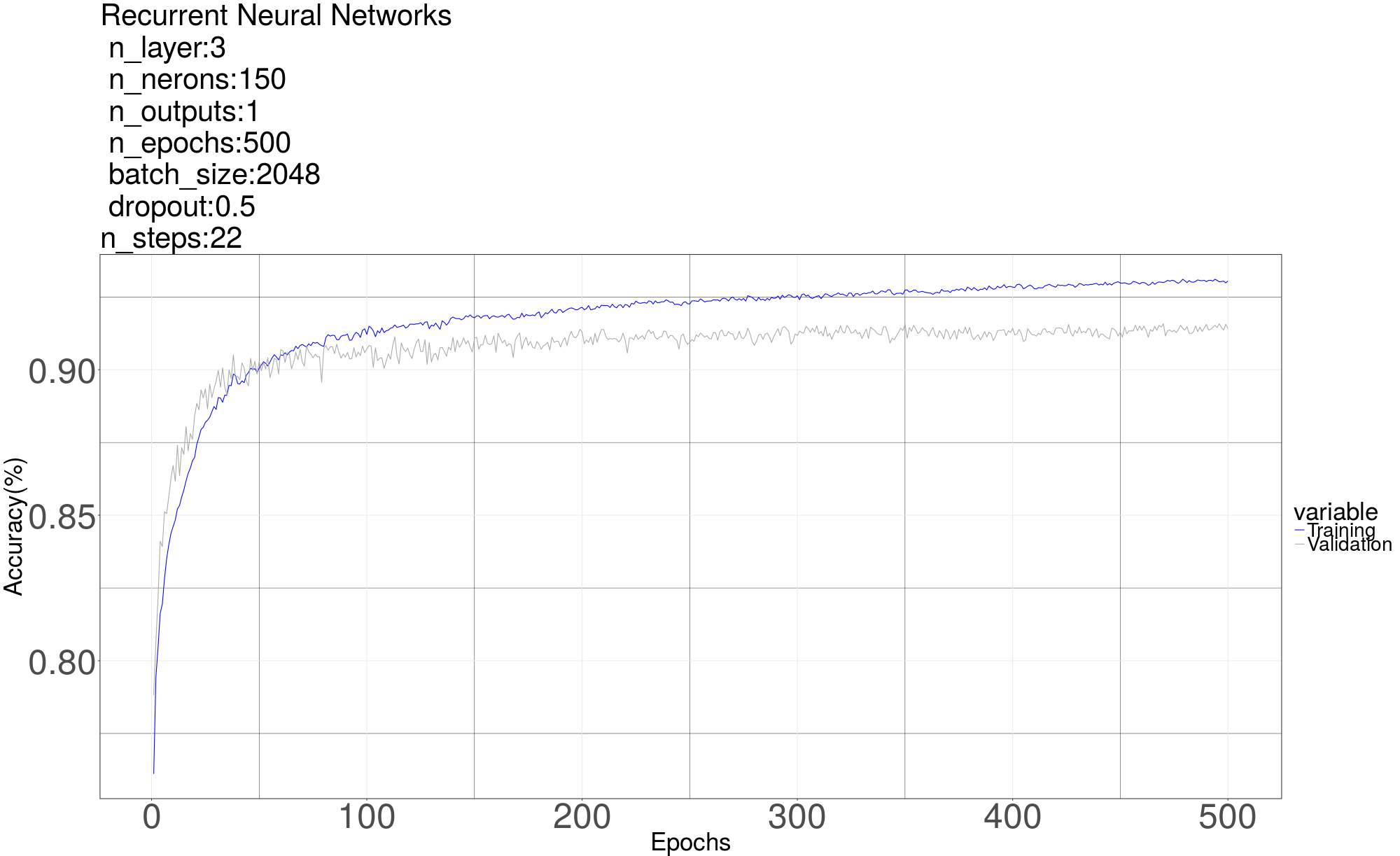}
\caption{\label{fig:RNN_4}Results of Recurrent Neural Network on Training and Validation set. Title contains information about configuration of the network.}
\end{figure}

\begin{figure}  
\centering
\includegraphics[width=1.0\textwidth]{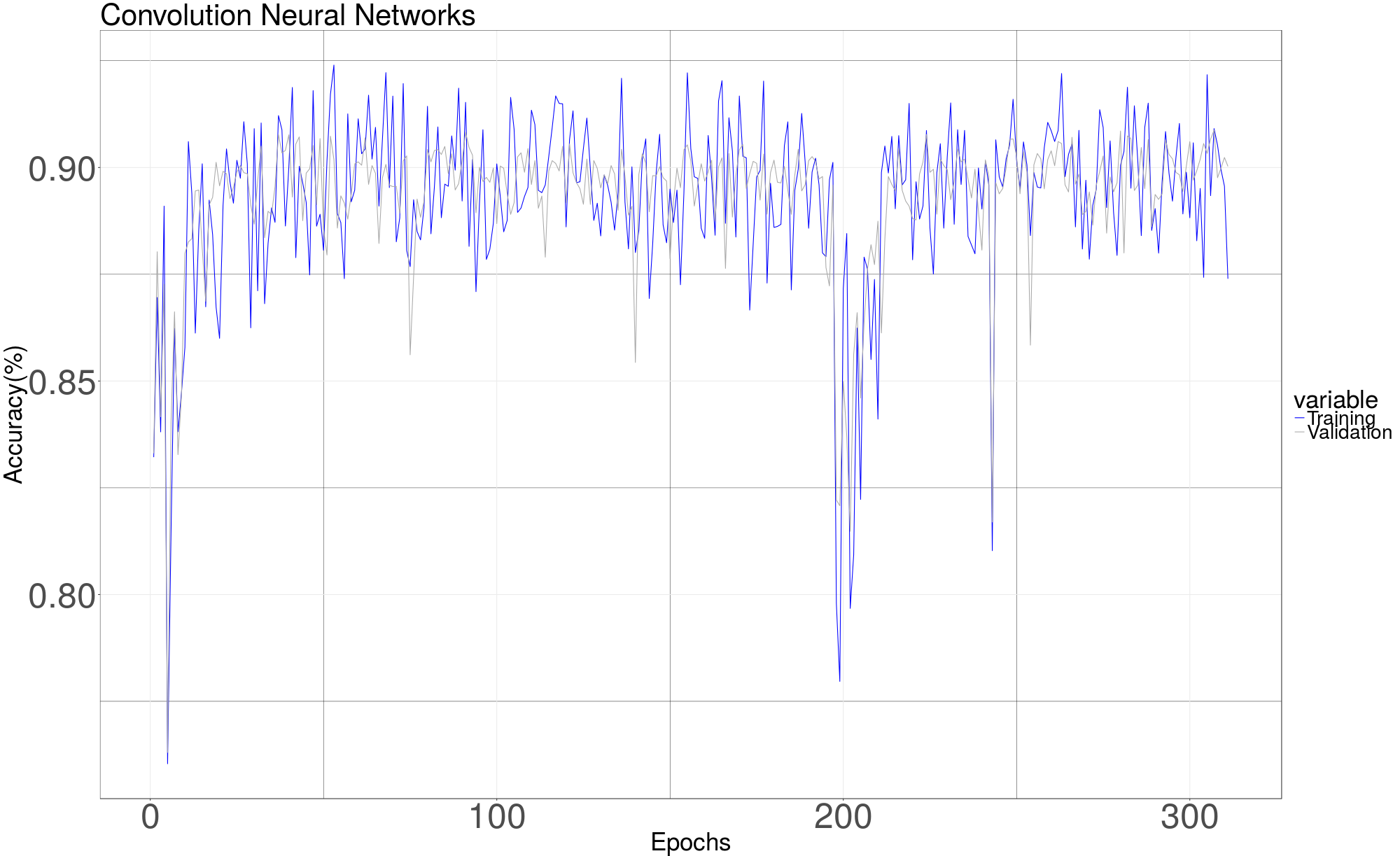}
\caption{\label{fig:CNN_1}Results of Convolution Neural Network on Training and Validation set. Title contains information about configuration of the network.}
\end{figure}

\bibliographystyle{IEEEtran}
\bibliography{sample}


\end{document}